\documentclass[a4paper]{jpconf}

\usepackage{graphicx}
\usepackage{hyperref,hypernat}
\usepackage{psfrag}
\usepackage{color}

\usepackage{amsmath,amssymb}

\newcommand{\ka}{{\kappa}}
\newcommand{\kk}{{\ka_{\rm K}}}

\newcommand{\Dt}{{D_{\rm 1}}}

\newcommand{\run}{\Delta_{\rm H}}

\newcommand{\om}{\omega}
\newcommand{\omm}{\omega_{\rm max}}
\newcommand{\ommin}{\omega_{\rm min}}

\newcommand{\Th}{{T_{\rm H}}}

\begin{document}

%\title{Phonon spectra with asymmetric sonic horizons}
\title{On the robustness of acoustic black hole spectra}

% \date{\today}

\author{S Finazzi$^1$ and R Parentani$^2$}
\address{$^1$ SISSA, via Bonomea 265, Trieste 34151, Italy; INFN sezione di Trieste}
\address{$^2$ Laboratoire de Physique Th\'eorique, CNRS UMR 8627, B\^at. 210, Universit\'e Paris-Sud 11, 91405 Orsay Cedex, France}

\ead{finazzi@sissa.it and renaud.parentani@th.u-psud.fr}

% \author[sissa]{
% \ead{liberati@sissa.it}
% 
% \author[iaa]{Carlos Barcel\'o}
% \ead{carlos@iaa.es}
% 
% 
% \address[sissa]{SISSA, via Beirut 2-4, Trieste 34151, Italy\\and INFN sezione di Trieste}
% % \address[infn]{INFN sezione di Trieste, via Valerio 2, 34127 Trieste, Italy}
% \address[iaa]{Instituto de Astrof\'{\i}sica de Andaluc\'{\i}a, CSIC, Camino Bajo de Hu\'etor 50, 18008 Granada, Spain}

% \cortext[cor1]{Corresponding author}

\begin{abstract}

We study the robustness of the %thermal 
%properties of the %of 
%phonons 
spectrum 
%spontaneously 
emitted by an acoustic black hole by considering %R3 
series of stationary flows 
% where the sonic horizon disappears as the flow 
%becomes either subsonic or supersonic. 
that %tends to 
become %everywhere 
either subsonic or supersonic, i.e.  when the %sonic 
horizon disappears.
We work with the superluminal Bogoliubov dispersion %in the  context 
of Bose--Einstein condensates.
We find that %to high accuracy 
the spectrum remains remarkably %is almost always 
Planckian until the %sonic 
horizon disappears.
%. In fact thermality is lost (in a continuous manner) 
%only in limiting cases
% tends to become everywhere either subsonic or supersonic. 
When the flow 
is everywhere supersonic, new pair creation %emission 
channels open.
%because the dimensionality of the stationary modes increases from three to four.
This will be the subject of a forthcoming work.

\end{abstract}

%----------------------------
\section{Introduction}
%---------------------------

In the hydrodynamic approximation, i.e. for long wavelengths, the propagation of sound waves in a moving fuid  is analogous to that of light in a curved spacetime~\cite{Unruh81,lr}. 
%It is therefore possible to define an 
More precisely, a %Rf an a must be used because one sounds like wouan ... 
one dimensional flow defines the %$1+1$
acoustic metric %geometry %defined  % by the following metric:
\begin{equation}\label{eq:metric}
 ds^2=-c^2 dt^2+(dx-v \, dt)^2~,
\end{equation}
where  $v$ is the flow velocity and $c$ the speed of sound.
Assuming that the fluid flows from right to left ($v<0$), a sonic horizon is present where $w = c+v$ crosses $0$.
This situation describes %mimics 
a black (white) hole horizon when $\kk \equiv \partial_x w $
evaluated at $w=0$ is positive (negative). %, and a white one when it is negative. %$\partial_x w < 0$.
One thus expects that such system would emit a Hawking flux, i.e. a thermal flux of phonons
at a temperature given by $\Th = \kk/2\pi$ (in units where $k_B= \hbar = 1$).

However, since Hawking radiation relies on short wavelength modes~\cite{TJ,Primer},
%could you melt in a single ref, the 91 and 93 TJpaper ?, if not leave it.
the dispersion of sound waves, which is neglected in the hydrodynamic approximation,
must be taken into account. % when computing the spectrum.
% emitted by an acoustic black hole. 
%To this end, Unruh wrote a dispersive wave equation in a supersonic flow~\cite{Unruh95}.
%Through a numerical analysis, he then showed that the spectrum was robust, provided the dispersive scale $\xi$, the ``healing length'', is much smaller than the surface gravity scale $1/\kappa$ which fixes the Hawking temperature $\Th = \kappa/2\pi$.
% (in units where $\hbar = k_{\rm B} = c = 1$).
%This insensitivity was then confirmed by analytical~\cite{BMPS95,Corley97,Tanaka99,UnruhSchu05} and numerical~\cite{CJ96,UnruhTrieste,MacherRP1,MacherBEC} methods in various physical system.
Following the original work~\cite{Unruh95}, this was done using analytical~\cite{BMPS95,Corley97,Tanaka99,UnruhSchu05} and numerical~\cite{CJ96,UnruhTrieste,MacherRP1,MacherBEC} methods. 
Provided the dispersive scale $\xi$, the ``healing length'', is much smaller than the surface gravity scale 
%R I put a c
$c/\kappa$, %In these works, it  %in various physical systems
it was found that the thermal properties of the spectrum are extremely robust. % which fixes the Hawking temperature $\Th = \kappa/2\pi$ (in units where $\hbar = 1$).
In fact, so robust that it is very difficult to characterize and analyze the small
{%leading 
deviations} with respect to the standard flux.
%R3

%However it turned out that the  are difficult %much harder 
%to characterize, and %so far
%there is no consensus on what are the %relevant 
%parameters that govern them. %the leading deviations. 

Our aim is to complete the analysis %Using this framework, it was shown
of~\cite{MacherRP1,MacherBEC} %Rf add MacherRP1,
where it was shown that the spectrum remains %remarkably 
Planckian
%to a high accuracy 
whenever % dispersive effects are sufficiently small.
$\Th \lesssim \omm/15$, where $\omm$ is a critical frequency
which scales as %the dispersive scale
$1/\xi$ but, {more importantly}, also depends on %$w_{\rm as} = 
$w(x=-\infty)$, 
the asymptotic value of
%velocity excess 
$c+v$. 
%For $\om > \omm$, %Above that critical frequency 
%the phonon spectrum vanishes because the negative norm modes, which are responsible 
%for the black hole radiation, no longer exist. 
%However, 
%Yet the results of~\cite{MacherBEC} were 
Since this result was obtained using %a very
symmetrical profiles like $w \sim D \tanh(\kappa x/D)$,
one may %thus 
wonder whether the robustness of the %Planckian 
spectrum was
not partially due to the symmetry of the flow w.r.t. the horizon, namely $w(-x) = - w(x)$.
%In this paper we try t
To investigate this question, % in this paper 
we shall consider flows which are highly %very 
asymmetric. % see FigXXX, 
%Furthermore, 
%R3
In particular we shall study the spectrum in
the interesting limiting cases % profiles and studying the limit 
%in which 
where the %sonic 
horizon disappears because
the flux becomes either %completely 
sub or supersonic. In parallel with this work,
we investigated in~\cite{smearhor} %other types of 
%locally modified
other classes of profiles %that %also 
generalizing %Rf
those of~\cite{MacherRP1,MacherBEC}. %Rf add MacherRP1, 
%R3

%In the same 

%flux of emitted quasi-particles vanishes when $\om$ approaches a certain critical frequency $\omm$, depending on the dispersive scale $\xi$ and on the asymptotic flow. However,  ($\omm/\kappa\gtrsim2.5$), the spectrum is expected to be Planckian, below the cutoff frequency $\omm$.
%However, the results of~\cite{MacherBEC} were obtained using a very symmetrical velocity profile.
% , where the absolute value of the difference between $c$ and $v$ in the two (supersonic and subsonic) asymptotic regions was equal.

%------------------------------------------------------
\section{Phonon spectra in asymmetric flows} %Settings}
%------------------------------------------------------
%
\begin{figure}
 \centering
 \psfrag{w}{\small $w(x)/c_0$}
 \psfrag{x}{\small $\ka x/c_0$}
 \includegraphics[width=0.6\textwidth]{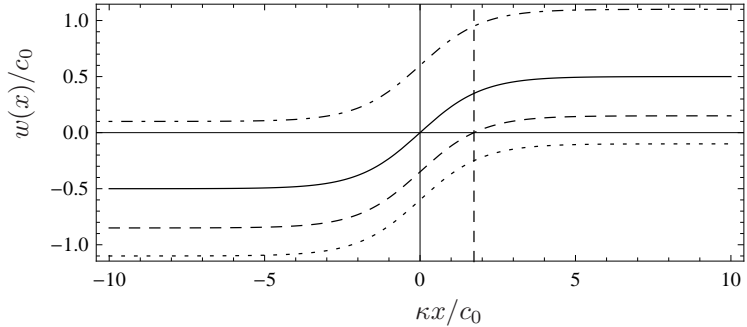}
 \caption{\label{fig:profile} %Three
Four velocity profiles $w=c+v$ of Eq. (\ref{vdparam}) with the same height $D $ %R3 ok ? c_0/\ka=0.5$ 
but
four % different
values of %the parameter 
$\Dt$. For $ \Dt= 0$ (solid line), one recovers the symmetric case of~\cite{MacherRP1}
with the Killing horizon (solid vertical line) at $x=0$.
For $-D<\Dt<0$ (dashed line), the horizon (dashed vertical line) is shifted to $x>0$.
In the last two cases, there is no horizon: for $\Dt<-D$ (dotted line), the flow is everywhere supersonic,
and for $\Dt>D$ (dotdashed line), it is subsonic.}
\end{figure}

We consider elongated condensates, stationary flowing along the longitudinal direction $x$. We assume that the transverse dimensions are small enough that the relevant phonon excitations are longitudinal. 
The system is thus effectively described by the $1+1$ dimension 
metric~\eqref{eq:metric} where $v$ and $c$ depend only on $x$.
%Following
%Unlike~\cite{MacherBEC}
We work with flows $w = v+c$ given by %fix the metric by choosing
%
%\begin{eqnarray}\label{eq:cv}
% c(x)&=&\chor+(1-q) w(x) >0,\nonumber \\
% v(x)&=&-\chor+q \, w(x) <0,
%\end{eqnarray}
%
%where $q$ is a constant and $w\equiv c+v$ is
% which specifies how $c+v$ is shared between $c$ and $v$. 
%
\begin{equation}
\frac{w(x)}{c_0 } = \Dt + D \, \tanh\left(\frac{\ka x}{D c_0}\right).
\label{vdparam}
\end{equation}
%
%The parameter
$D$ %, ranging from 0 to 1, Needed ?? Not really
%R3 determines the size of the region where $w$ varies linearly in $x$, 
fixes the gap $[w(\infty) - w(-\infty)]/2 c_0$, 
while $\Dt$, which was taken to be $0$ in~\cite{MacherRP1,MacherBEC}, %Rf add MacherRP1,
fixes the asymmetry between sub and supersonic regimes. When $\Dt > D$, the flow is everywhere subsonic, and when $\Dt <- D $, it is %everywhere 
supersonic, see Fig.~\ref{fig:profile}. 
Instead, when $|\Dt|<D$, there is 
%there are sub and a supersonic %and subsonic region that are separated by 
a sonic (Killing) horizon where $w=0$, localized at
\begin{equation}
 x_{\rm H}=-\frac{Dc_0}{\ka}\,\mbox{arctanh}\left(\frac{\Dt}{D}\right).
\end{equation}
It separates the sub ($x>x_{\rm H}$) from the supersonic region ($x < x_{\rm H}$).
In such a flow, %this regime, 
when ignoring %short distance
dispersion, the spectrum of upstream phonons spontaneously emitted from the horizon
would be very simple, and
would strictly correspond to the Hawking radiation~\cite{Unruh81,Unruh95}: %R08
It would follow a Planck law at Hawking temperature $\Th={\kk}/{2\pi}$. %, in units where $\hbar = k_B = 1$. 
The surface gravity $\kk$ is here
\begin{equation}\label{eq:temprel}
%\Th=\frac{\kk}{2\pi} \equiv \frac{1}{2\pi}
\kk  \equiv 
\partial_x (c+v) \vert_{x = x_H} %R Stefano not 0 ?????? % = \ka + \kt,
= \ka % \frac{\ka}{2\pi} 
\left[1-\left(\frac{\Dt}{D}\right)^{2}\right]. % < \ka. %\frac{\ka}{2\pi}, %R   
\end{equation}
%
%However, in the presence of a dispersive background, 
Taking dispersion into account, the spectrum becomes much more complicated.
%one obtains Bogoliubov dispersion
% of Eq. (\ref{eq:dispersion_hom}), 
In this paper, we consider %shall compute the spectrum %address this question by considering %focusing on 
phonons in Bose--Einstein condensates (BEC) with the Bogoliubov %, where the 
dispersion  %of propagating perturbations 
\begin{equation}\label{eq:dispersion_hom}
%\Omega^2(k) = %
(\om-vk)^2=\Omega^2(k)=
c^2 k^2 + \frac{\hbar^2 k^4}{4 m^2 } = c^2 k^2\left( 1 +  \frac{\xi^2 k^2}{2}\right),
\end{equation}
where $\om$ is the lab conserved frequency.
%the phonon spectrum is modified. %~\cite{MacherBEC}. 
In the sequel, %next section, 
following the numerical procedure of~\cite{MacherBEC}, % where details can be found
% , Using a modified version of the code written for~\cite{MacherBEC}, 
we study the deviations of $n_\om$, the phonon spectrum,  w.r.t. the standard Planck one,
by varying the offset parameter $\Dt$ while keeping fix all the other parameters. %%R OK ? %S Yes

%spectral properties of the mean flux %occupation number 
%$n_\om$ of phonons emitted to the right. % of the horizon and its from Planckianity.

%-------------------------------------------
%\section{Results}
%-------------------------------------------

\begin{figure}
\includegraphics[width=.49\textwidth]{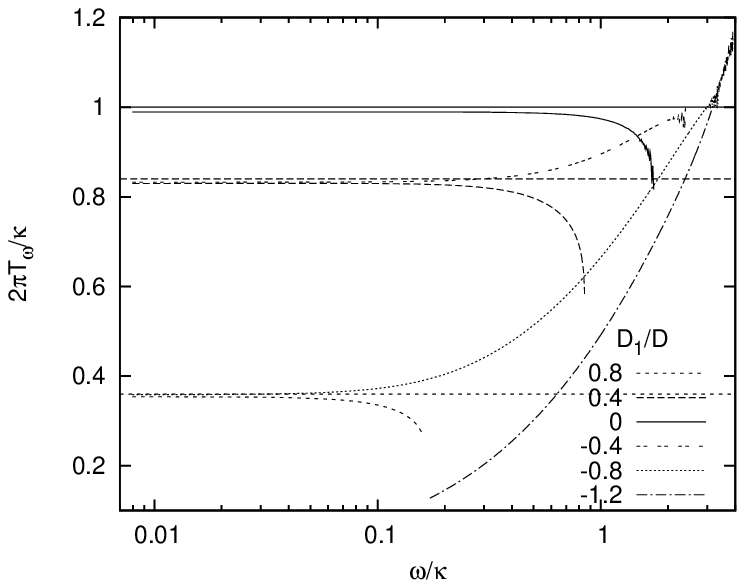}
\includegraphics[width=.49\textwidth]{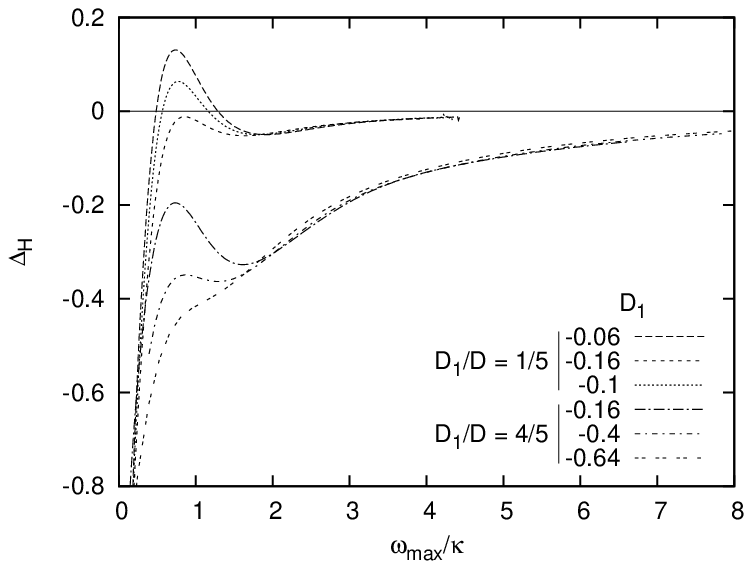}
\caption{\label{fig:tom}Left panel: $T_\om$ for %R3 various 
6 values of $\Dt/D$,
with $D=0.5$ and $\kappa\xi/\sqrt{2}c=0.1$ fixed. The horizontal lines represent $\Th(\Dt)$ of Eq.~\eqref{eq:temprel}, respectively for $|\Dt|=0,0.2,0.4$.
% The spectrum is almost Planckian for $\om<\omm$ at temperature $\Th$ when $|\Dt|<D$. When $\Dt\to -D$ the deviation from Planckianity becose more relevant. When $\Dt<-D$ and the flux is everywhere supersonic, no horizon is present and the spectrum is not Planckian even at low frequency.
For $\Dt<-D$ the curve is truncated at $\ommin$ as explained in the text.
%because the standard analysis does not apply and further techniques are required~\cite{4x4}.
Right panel: $\run$ as a function of $\omm/\kappa$ for two values of $\Dt/D$ and  various $\Dt$.
When $\omm/\kappa$ is large enough, deviations from Planckianity are small and, more importantly, 
governed by the ratio $\Dt/D$ only.}
\end{figure}

To characterize these deviations, %of the phonon spectrum 
%mean occupation phonon number $n_\om$ of phonons,
% in place of $n_\om$ 
we use the temperature function $T_\om$ defined by
\begin{equation}\label{eq:Tom}
  n_\om \equiv \frac{1}{\exp(%\hbar \om/k_B 
\om/ T_\om)-1},
 \end{equation}
which is %presents the great advantage of being 
constant when the spectrum is Planckian. This quantity thus allows a direct comparison with the 
%predicted 
Hawking
temperature $\Th=\kk/2\pi$ of Eq.~\eqref{eq:temprel}. % of \eqref{eq:temprel}.
In Fig.~\ref{fig:tom}, $T_\om$ and $\Th$ (horizontal lines)
are plotted %as functions of $\om$
for various values of $\Dt$.
% together with %the Hawking temperature 
%$\Th=\kk/2\pi$ (horizontal lines) of Eq.~\eqref{eq:temprel}.
We distinguish four regimes:
\begin{itemize}
 \item For $|\Dt|$ sufficiently smaller than $D$, 
the spectrum is %R3 remains 
Planckian until $\om$ approaches $\omm$, where it vanishes.
%since $T_\om$ hardly depends on $\om$.
For $\Dt > 0$, $\omm$ decreases %(increases)
since $\vert w(x = - \infty)\vert$ does so, and conversely for $\Dt < 0$. %(increases).
%depends on the sign of $\Dt$ since it affects the asymptotic value of $w$.
However for both signs, %Moreover 
at low frequency $T_\om$ %actual temperature 
closely follows %the relativistic value 
$\kk/2\pi$ of \eqref{eq:temprel}.%R
%$\Th$ which is even in $\Dt$, see   
%
 \item When $|\Dt|\to D$, $\kk$ of \eqref{eq:temprel} drops down to $0$ and the sonic horizon disappears. 
%R3
In this critical regime, deviations from Planckianity appears even at low frequency.
 \item For $\Dt>D$, the flow is everywhere subsonic. 
There is no particle production because there are no negative norm modes with $\om > 0$.

 \item For $\Dt< - D$, the flow is supersonic. A new critical frequency $\ommin<\omm$ appears.
For $0< \om<\ommin$, there are now $4$ asymptotic in and out modes, and
%. Therefore 
the scattering matrix %particle production 
is  %described by a 
$4\times4$.
% Bogouliubov matrix, while the present analysis is restricted 
Since the code of~\cite{MacherBEC} is designed to the $3\times3$ case, this regime
requires %a to write %R3
a new code~\cite{4x4}.
However, for $\ommin<\om<\omm$ there are only 3 modes, and our code can handle this frequency band.
%our treatment can be applied. 
In this range, as can be seen in the left panel, $n_\om$ is close to the case where $\Dt > - D$, even though
%However $\kk$ is not defined, as well as the Haking temperature $\Th$. As a consequence 
the spectrum is not at all Planckian. % and $T_\om$ is not constant. 
\end{itemize}
%R Stefano, could you please replace the p by a 1 in the subsc of D in the caption ?
%S Ok. Thanks
% could you please as well give the values of \Dt/D so that larger/smaller than 1 is easily seen ? Do you agree ?
%S Yes, it is a good idea

Furthermore, we checked that the {\it shape} of the spectrum is hardly changed %R3
when %i.e. by 
varying the healing length $\xi$. %R3  is varied. 
Namely, spectra calculated with different values of $\xi$ almost coincide when $T_\om$ is plotted versus the rescaled quantity $\om/\omm(\xi)$.
%R3
%
To identify %R3 what are 
the parameters governing the deviations from Planckianity we define
\begin{equation}
 \run\equiv\frac{f(\omega=\Th)-f_{\Th}(\omega=\Th)}{f_{\Th}(\omega=\Th)},
\end{equation}
where $f(\om)\equiv\om n_\om$ is the actual energy flux %of the emitted particles 
and $f_{\Th}(\om)$ is the thermal flux at temperature $\Th$. 
In Fig.~\ref{fig:tom}, right panel, $\run$ is plotted as a function of $\omm(\xi)/\kappa$. 
%R3 already defined.  
Various values of $\Dt$ are used, while the ratio $\Dt/D$ is kept constant respectively for
the three lower curves ($\Dt/D=-1/5$), and for the three upper curves ($\Dt/D=4/5$).
As expected, % from previous analysis~\cite{MacherBEC}, 
$\run$ goes to 0
% the difference of the spectrum from a Planckian one goes to $0$ 
for large values of $\omm/\kappa$. What we learn here is that %In addition, for $\omm/\ka\gtrsim 2$,
the leading deviations from Planckianity do not depend	 separately on $\Dt$ or $D$ but only on their ratio.

% When $\Dt

%------------------------------------------
\section{Comments and conclusions}\label{sec:conclusion}
%------------------------------------------

In this work, we %R3 further
% investigate 
have studied the robustness of the properties of black hole radiation in BEC
by considering flows characterized by a high asymmetry between the sub and the supersonic region.
We found that the spectrum %R3 at low frequencies 
remains Planckian %with high accuracy unless 
until one approaches the critical cases ($\vert \Dt \vert \to D$) 
where the surface gravity \eqref{eq:temprel} vanishes. 
% Only when reaching the limit in which either the sub- or the super-sonic region disappears, the spectrum is heavly modified.

In fact, %However, 
for $0 < \Dt \to D$,
since the fluid flows in the supersonic region with a velocity just above that %close to the speed 
of sound, the cutoff frequency $\omm$ %also %R3
goes to zero.
Hence the range of frequencies where the temperature %function 
$T_\om$ is nearly constant shrinks and % finally
disappears in the limit. As expected, %where there is no horizon, % by the absence of the horizon, 
no emission of particles is found when the flow remains everywhere subsonic.

The situation is much more interesting in the opposite case, for $0 > \Dt \to - D$,
when pushing the velocity of the subsonic region towards the speed of sound. In this regime, even though
$\kk$ of
\eqref{eq:temprel} goes again %R3
to $0$, $\omm$ now increases. As a result, the low frequency plateau of $T_\om$ at low temperature again shrinks, but
the spectrum becomes more and more blue. This can be understood from the fact that 
%for $v\approx c$, as the sonic horizon moves into the right asymptotic region. As a consequence, 
more dispersive modes, 
i.e. %high frequency 
modes with higher $\om$, have still their turning point near $x= 0$
%which can more deeply penetrate inside the supersonic region, feel a larger effective surface gravity with respect to the low frequency ones. 
where the gradient of $w$ is $\kappa %$, which is now much higher than $
\gg \kk$.
%t the relativisticsurface gravity.
%Accordingly, the temperature function initially increases with frequency before vanishing for $\omega\to\omm$. 

When %R3 %crossing the limit 
$\Dt < - D$, %and considering %When the 
there is no sonic horizon since the flow is everywhere supersonic. 
Yet the spectrum is continuously deformed even though the Planck character is completely lost in
the frequency range $\ommin < \om < \omm$, where $\ommin$ is a new critical frequency. 
For $\om<\ommin$, new scattering channels open because there is a fourth asymptotic mode. 
Investigations of %R in ?
this case~\cite{4x4} are in progress.

To conclude, we stress that the thermal spectrum at the na\"ive surface gravity~\eqref{eq:temprel} 
provides a reliable approximation of the actual phonon spectrum in all cases but the critical ones where the sonic horizon disappears. %only far from limit situations. 
In these extreme cases, the spectrum is no longer thermal and its properties are %R3 rather %and the spectrum are 
governed by the dispersive properties of field. 
% of fferent penetration length beyond the horizon as a function of frequency and averaging effects related to dispersion~\cite{smearhor}.

%----------------------------
\section*{Acknowledgments}
%----------------------------

We are grateful to A. Coutant and S. Liberati for interesting %R3 many 
discussions. % that took place during this project. 
%We are also grateful to Jean Macher for useful explanations concerning the code he wrote for~\cite{MacherBEC}. 

\section*{References}
\bibliographystyle{iopart-num}

\begin{thebibliography}{10}
\expandafter\ifx\csname url\endcsname\relax
  \def\url#1{{\tt #1}}\fi
\expandafter\ifx\csname urlprefix\endcsname\relax\def\urlprefix{URL }\fi
\providecommand{\eprint}[2][]{\url{#2}}

\bibitem{Unruh81}
Unruh W~G 1981 {\em Phys. Rev. Lett.\/} {\bf 46} 1351--1353

\bibitem{lr}
Barcelo C, Liberati S and Visser M 2005 {\em Living Rev. Rel.\/} {\bf 8} 12

\bibitem{TJ}
Jacobson T 1991 {\em Phys. Rev. D\/} {\bf 44} 1731--1739; % \bibitem{93}
% Jacobson T 
1993 {\em Phys. Rev. D\/} {\bf 48} 728--741

\bibitem{Primer}
Brout R, Massar S, Parentani R and Spindel P 1995 {\em Phys. Rept.\/} {\bf 260}
  329--454

\bibitem{Unruh95}
Unruh W~G 1995 {\em Phys. Rev. D\/} {\bf 51} 2827--2838

\bibitem{BMPS95}
Brout R, Massar S, Parentani R and Spindel P 1995 {\em Phys. Rev. D\/} {\bf 52}
  4559--4568

\bibitem{Corley97}
Corley S 1998 {\em Phys. Rev. D\/} {\bf 57} 6280--6291

\bibitem{Tanaka99}
Himemoto Y and Tanaka T 2000 {\em Phys. Rev. D\/} {\bf 61} 064004

\bibitem{UnruhSchu05}
Unruh W~G and Schutzhold R 2005 {\em Phys. Rev. D\/} {\bf 71} 024028

\bibitem{CJ96}
Corley S and Jacobson T 1996 {\em Phys. Rev. D\/} {\bf 54} 1568--1586

\bibitem{UnruhTrieste}
Unruh W~G 2007 {\em PoS\/} {\bf QG-PH} 039

\bibitem{MacherRP1}
Macher J and Parentani R 2009 {\em Phys. Rev. D\/} {\bf 79} 124008

\bibitem{MacherBEC}
{Macher} J and {Parentani} R 2009 {\em Phys. Rev. A\/} {\bf 80} 043601

\bibitem{smearhor}
{Finazzi} S and {Parentani} R 2010  (\textit{Preprint} \eprint{arXiv:1012.1556
  [gr-qc]})

\bibitem{4x4}
Coutant A, Finazzi S, Liberati S and Parentani R {\em in preparation\/}

\end{thebibliography}

\providecommand{\newblock}{}

\end{document}